\documentclass[10pt]{article}
\usepackage{graphics,epsfig}
\usepackage[]{graphicx}
\usepackage{latexsym}

\begin{document}

\title{Scopes and limits of modality in quantum mechanics}

\author{{\sc Graciela
Domenech}\thanks{%
Fellow of the Consejo Nacional de Investigaciones Cient\'{\i}ficas
y T\'ecnicas (CONICET)} $^{1, 3}$,\  \  {\sc Hector Freytes}
$^{2}$\ {\sc and} \ {\sc Christian de Ronde} $^{3, 4}$}

\maketitle

\begin{center}

\begin{small}
1. Instituto de Astronom\'{\i}a y F\'{\i}sica del Espacio (IAFE)\\
Casilla de Correo 67, Sucursal 28, 1428 Buenos Aires, Argentina\\
2. Dipartimento di Scienze e Pedagogiche e Filosofiche -
Universita degli Studi di Cagliari \\ Via Is Mirrionis 1, 09123,
Cagliari - Italia\\ 3. Center Leo Apostel (CLEA)\\ 4. Foundations
of the Exact Sciences (FUND) \\ Brussels Free University -
Krijgskundestraat 33, 1160 Brussels - Belgium
\end{small}
\end{center}

\begin{abstract}

We develop an algebraic frame for the simultaneous treatment of
actual and possible properties of quantum systems. We show that,
in spite of the fact that the language is enriched with the
addition of a modal operator to the orthomodular structure,
contextuality remains a central feature of quantum systems.

\end{abstract}

\maketitle

\bibliography{pom}

\newtheorem{theo}{Theorem}[section]

\newtheorem{definition}[theo]{Definition}

\newtheorem{lem}[theo]{Lemma}

\newtheorem{prop}[theo]{Proposition}

\newtheorem{coro}[theo]{Corollary}

\newtheorem{exam}[theo]{Example}

\newtheorem{rema}[theo]{Remark}{\hspace*{4mm}}

\newtheorem{example}[theo]{Example}

\newcommand{\proof}{\noindent {\em Proof:\/}{\hspace*{4mm}}}

\newcommand{\qed}{\hfill$\Box$}

\section{Introduction}

Contextuality is one of the main features of the discourse about
quantum systems and has been studied from different approaches. We
are interested here in algebraic versions related to partial
valuations of the orthomodular lattice of closed subspaces of
Hilbert space. This proposal allows to identify  the constraints
imposed by the structure to the relation between actuality and
possibility and the discourse that includes both type of
propositions.

The paper is organized as follows: in Sec. 2 we discuss the
contextual character of quantum systems from an algebraic
perspective. Sec. 3 introduces the desiderata of modal
interpretations of quantum mechanics involved in our treatment. We
devote Sec. 4 to expose the algebraic structure which represents
the orthomodular lattice enriched with modal operators. In Sec. 5
we show how the discourse about properties is genuinely enlarged
giving an adequate framework to represent the Born rule. In Sec. 6
we use Kochen-Specker theorem to prove that the contextual
character of quantum mechanics is maintained even when the
discourse is enriched with  modalities. Finally, we outline our
conclusions.

\section{Contextuality in quantum systems}

In paper \cite{DF}, we dealt with the problem of the limits
imposed by the orthomodular structure of projection operators to
the possibility of thinking of properties possessed by an isolated
quantum system. This is an important point in the discussion about
quantum systems because almost every problem in the relation
between the mathematical formalism and what may be called our
experience about the behavior of physical objects can be encoded
in the question about the possible meaning of the proposition {\it
the physical magnitude ${\cal A}$ has a value and the value is
this or that real number}. Already from the first formalizations,
this point was recognized. For example, P.A.M. Dirac stated in his
famous book: ``The expression that an observable `has a particular
value' for a particular state is permissible in quantum mechanics
in the special case when a measurement of the observable is
certain to lead to the particular value, so that the state is an
eigenstate of the observable. It may easily be verified from the
algebra that, with this restricted meaning for an observable
`having a value', if two observables have values for a particular
state, then for this state the sum of the two observables (if the
sum is an observable) has a value equal to the sum of the values
of the two observables separately and the product of the two
observables (if this product is an observable) has a value equal
to the product of the values of the two observables separately''
\cite{Dirac}. This last point is the requirement of the functional
compatibility condition (FUNC), to which we will return later. As
long as we limit ourselves to speaking about measuring results and
avoid being concerned with what happens to nature when she is not
measured, quantum mechanics carries out predictions with great
accuracy. But if we naively try to interpret eigenvalues as actual
values of the physical properties of a system, we are faced with
all kind of no-go theorems that preclude this possibility. Most
remarkable is the Kochen-Specker (KS) theorem that rules out the
non-contextual assignment of values to physical magnitudes
\cite{ks}. In \cite{DF} we gave algebraic, topological and
categorial versions of the KS theorem. To include modalities in
our analysis we will deal with the algebraic form in terms of
valuations. So we recall the main features of our discussion
there.

Let ${\mathcal H}$ be the Hilbert space associated with the
physical system and $L({\mathcal H})$ be the set of closed
subspaces on ${\mathcal H}$. If we consider the set of these
subspaces ordered by inclusion, then $L({\mathcal H})$ is a
complete orthomodular lattice \cite{MM}. It is well known that
each self-adjoint operator $\bf A$ representing a physical
magnitude ${\cal A}$ has associated a Boolean sublattice $W_A$ of
$L({\mathcal H})$. More precisely, $W_A$ is the Boolean algebra of
projectors ${\bf P}_i$ of the spectral decomposition ${\bf
A}=\sum_{i} a_i {\bf P}_i$. We will refer to $W_A$ as the spectral
algebra of the operator $\bf A$. Any proposition about the system
is represented by an element of $L({\mathcal H})$ which is the
algebra of quantum logic introduced by G. Birkhoff and J. von
Neumann \cite{BV}.

Assigning values to a physical quantity ${\cal A}$ is equivalent
to establishing a Boolean homomorphism $v: W_A \rightarrow {\bf
2}$ \cite{IB}, being ${\bf 2}$ the two elements Boolean algebra.
So it is natural to consider the following definition:

\begin{definition}

Let $(W_i)_{i\in I}$ be the family of Boolean sublattices of
$L({\mathcal H})$. A global valuation over $L({\mathcal H})$ is a
family of Boolean homomorphisms $(v_i: W_i \rightarrow {\bf
2})_{i\in I}$ such that $v_i\mid W_i \cap W_j = v_j\mid W_i \cap
W_j$ for each $i,j \in I$

\end{definition}

Were it possible, this global valuation would give the values of
all magnitudes at the same time maintaining a {\it compatibility
condition} in the sense that whenever two magnitudes share one or
more projectors, the values assigned to those projectors are the
same from every context.

But the KS theorem assures that we cannot assign real numbers
pertaining to their spectra to operators ${\bf A}$ in such a way
to satisfy the functional composition principle (FUNC) which is
the expression of the ``natural'' requirement mentioned by Dirac
that, for any operator ${\bf A}$ representing a dynamical variable
and any real-valued function $f({\bf A})$, the value of $f({\bf
A})$ is the corresponding function of the value of ${\bf A}$.  In
the algebraic terms of the previous definition, the KS theorem
reads:

\begin{theo}\label{CS2}
If $\mathcal{H}$ is a Hilbert space such that $dim({\cal H}) > 2$,
then a global valuation over $L({\mathcal H})$ is not possible.
\qed
\\

\end{theo}

Of course contextual valuations allow us to refer to different
sets of actual properties of the system which define its state in
each case. Algebraically, a {\it contextual valuation} is a
Boolean valuation over one chosen spectral algebra. In classical
particle mechanics it is possible to define a Boolean valuation of
all propositions, that is to say, it is possible to give a value
to all the properties in such a way of satisfying FUNC. This
possibility is lost in the quantum case.

Now we intend to show how this discussion is able to include {\it
modalities}, i.e. to consider also possibility and necessity of
propositions about the properties of physical systems: what {\it
may be the case} (what the possible physical situations are) and
what {\it is necessarily the case}. The consideration of {\it
possibility} is of course always present in quantum theories. What
we propose here is an algebraic consideration of this type of
sentences and their articulation with those about actual
properties. Several attempts to obtain modal extensions of the
orthomodular systems are found in the literature. One possibility
developed in \cite{Dish} allows to embed orthomodular
propositional systems in modal systems. Another extension is
provided by adding quantifiers to the orthomodular structure
\cite{JAN1, JAN2}, so generalizing the monadic extension of the
Boolean algebras \cite{HAL}. In our case, we enrich the
orthomodular structure with a modal operator thus obtaining an
algebraic variety such that each orthomodular lattice can be
represented by an algebra of this variety. On the other hand, this
operator acts as a quantifier in the sense of \cite{JAN1, JAN2}.
The physical motivation for this construction is the purpose to
link consistently the propositions about actual and possible
properties of the system in a single structure.

\section{Modal interpretations}

Modal interpretations of quantum mechanics \cite{vF91, D88, D89,
DD} face the problem of finding an objective reading of the
accepted mathematical formalism of the theory, a reading ``in
terms of properties possessed by physical systems, independently
of consciousness and measurements (in the sense of human
interventions)'' \cite{DD}. These interpretations intend to
consistently include the possible properties of the system in the
discourse and so find a new link between the state of the system
and the probabilistic character of its properties, namely,
sustaining that the interpretation of the quantum state must
contain a modal aspect. The name {\it modal interpretation} was
for the first time used by B. van Fraassen \cite{BvF} following
{\it modal logic}, precisely the logic that deals with possibility
and necessity. The fundamental point is the purpose of
interpreting ``the formalism as providing information about
properties of physical systems'' \cite{DD}. In this context, a
physical property of a system is ``a definite value of a physical
quantity belonging to this system; i.e., a feature of physical
reality'' \cite{DD}. As usual, definite values of physical
magnitudes correspond to yes/no propositions represented by
orthogonal projection operators acting on vectors belonging to the
Hilbert space of the (pure) states of the system \cite{jauch}.

Modal interpretations may be thought to be a study of the
constraints under which one is able to talk a consistent classical
discourse without contradiction with the quantum formalism. To
study this issue and in order to avoid inconsistencies, we face
the problem of modalities in the frame of algebraic logic. To do
so, we build a variety that is an expansion of the orthomodular
lattices by adding an operator, {\it the possibility operator}, to
these structures. It will represent the possibility of occurrence
of a property, measurable in terms of the Born rule. The analysis
of the changes introduced by allowing modalities will be
performed, as in \cite{DF}, for the case of pure states. In spite
of the restrictions this imposes to the comparison with the
general case of modal interpretations, we think it contributes to
enlighten the discussion all the same. In a following step, we
will extend the treatment to the factorized space of subsystems.

\section{An algebraic study of modality}

First we recall from \cite{Bur, MM, BD} some notions of the
universal algebra and lattice theory that will play an important
role in what follows. For each algebra $A$, we denote by $Con(A)$,
the congruence lattice of $A$, the diagonal congruence is denoted
by $\Delta$ and the largest congruence $A^2$ is denoted by
$\nabla$. $\theta$ is called  {\it factor congruence} iff there is
a congruence $\theta^*$ on $A$ such that, $\theta \land \theta^* =
\Delta$, $\theta \lor \theta^* = \nabla$ and $\theta$ permutes
with $\theta^*$. If $\theta$ and $\theta^*$ is a pair of factor
congruences on $A$ then $A \cong A/\theta \times A/\theta^*$. $A$
is {\it directly indecomposable} if $A$ is not isomorphic to a
product of two non trivial algebras or, equivalently
$\Delta,\nabla$ are the only factor congruences in $A$. We say
that $A$ is {\it subdirect product} of a family of $(A_i)_{i\in
I}$ of algebras if there exists an embedding $f: A \rightarrow
\prod_{i\in I} A_i$ such that $\pi_i f : A\! \rightarrow A_i$ is a
surjective homomorphism for each $i\in I$ where $\pi_i$ is the
projection onto $A_i$. $A$ is {\it subdirectly irreducible} iff
$A$ is trivial or there is a minimum congruence in $Con(A) -
\Delta$. It is clear that a subdirectly irreducible algebra is
directly indecomposable. An important result due to Birkhoff is
that every algebra $A$ is subdirect product of subdirectly
irreducibles algebras. In a Boolean algebra $A$, congruences are
identifiable to certain subsets called {\it filters}. $F \subset
A$ is a filter iff it satisfies: if $a\in F$ and $a\leq x$ then
$x\in F$ and if $a,b\in F$ then $a\land b \in F$. $F$ is a proper
filter iff $F\not = A$ or, equivalently, $0\not \in F$. If
$X\subseteq A$, the filter $F_X$ generated by $X$ is the minimum
filter containing $X$. A proper filter $F$ is maximal iff the
quotient algebra $A/F$ is isomorphic to $\bf 2$. It is well known
that each proper filter can be extended to a maximal one.

We denote by ${\cal OML}$ the variety of orthomodular lattices.
Let $L=\langle L,\lor,\land, \neg, 0, 1\rangle$ be an orthomodular
lattice. Given $a, b, c$ in $L$, we write: $(a,b,c)D$\ \   iff
$(a\lor b)\land c = (a\land c)\lor (b\land c)$; $(a,b,c)D^{*}$ iff
$(a\land b)\lor c = (a\lor c)\land (b\lor c)$ and $(a,b,c)T$\ \
iff $(a,b,c)D$, (a,b,c)$D^{*}$ hold  for all permutations of $a,
b, c$. An element $z$ of a lattice $L$ is called a {\it central}
iff for all elements $a,b\in L$ we have\ $(a,b,z)T$. We denote by
$Z(L)$ the set of all central elements of $L$ and it is called the
{\it center} of $L$. $Z(L)$  is a Boolean sublattice of $L$ {\rm
\cite[Theorem 4.15]{MM}}.

\begin{prop}\label{eqcentro} {\rm \cite[Lemma 29.9 and Lemma
29.16]{MM}} Let $L$ be an orthomodular lattice then we have

\begin{enumerate}
\item
$z \in Z(L)$ if and only if $a = (a\land z) \lor (a \land \neg z)$
for each $a\in L$

\item
If $L$ is complete then $Z(L)$ is a complete lattice and for each
family $(z_i)_i$ in Z(L) and $a\in L$, $a \land \bigvee z_i =
\bigvee (a \land Z_i)$.
\end{enumerate}
\qed
\end{prop}

Factor congruences in $L$ are identifiable to the elements of the
center $Z(L)$. More precisely if $z\in Z(L)$, the binary relation
${\Theta}_z$ on $A$ defined by $a  \Theta_z b$  iff $a\land z =
b\land z$  is a congruence on $L$, such that $L\cong
L/{\Theta}_z\times L/{\Theta}_{\neg z}$.

\vspace{1cm}

Now we build up a framework to include modal propositions in the
same structure as actual ones. To do so, we enrich the
orthomodular lattice with a modal operator taking into account the
following considerations:

\begin{enumerate}

\item
Propositions about the properties of the physical system will be
interpreted in the orthomodular lattice of subspaces of the
Hilbert space of the (pure) states of the system. Thus we will
retain this structure in our extension.

\item
Given a proposition about the system, it is possible to define a
context from which one can predicate with certainty about it (and
about a set of propositions that are compatible with it) and
predicate probabilities about the other ones. This is to say that
one may predicate truth or falsity of all possibilities at the
same time, i.e. possibilities allow an interpretation in a Boolean
algebra. In rigorous terms, for each proposition $P$, if we refer
with $\Diamond P$ to the possibility of $P$, then $\Diamond P$
will be a central element of the orthomodular structure.

\item
If $P$ is a proposition about the system and $P$ occurs, then it
is trivially possible that $P$ occurs. This is expressed as $P
\leq \Diamond P$.

\item
To assume an actual property and a complete set of properties that
are compatible with it determines a context in which the classical
discourse holds. Classical consequences that are compatible with
it, for example probability asignements to the actuality of other
propositions, share the classical frame. These consequences are
the same ones as those which would be obtained by considering the
original actual property as a possible one. This is interpreted
as, if $P$ is a property of the system, $\Diamond P$ is the
smallest central element greater than $P$.

\end{enumerate}

The algebraic study will be performed using the necessity operator
$\Box$  instead of the possibility one $\Diamond$ because of
technical reasons. Then it will be possible to define the
possibility operator from the necessity one.

\begin{definition}
{\rm Let $A$ be an orthomodular lattice. We say that $A$ is {\it
Boolean saturated} iff for each $a\in A$ the set $\{z\in Z(A):
z\leq a \}$ has a maximum. In this case such maximum is denoted by
$\Box (a)$.

}
\end{definition}

\begin{example}
{\rm In view of Proposition \ref{eqcentro}, orthomodular complete
lattices with $\Box(a) = \bigvee \{z \in Z(L) : z \leq a \}$ as an
operator, are examples of Boolean saturated orthomodular
lattices.}
\end{example}

\begin{prop}\label{PROST}
Let $A$ be an orthomodular lattice. Then $A$ is Boolean saturated
iff there exists an unary operator $\Box$ acting on the elements
of $A$ satisfying

\begin{enumerate}

\item[S1]
$\Box x \leq x$

\item[S2]
$\Box 1 = 1$

\item[S3]
$\Box \Box x = \Box x$

\item[S4]
$\Box(x \land y) = \Box(x) \land \Box(y)$

\item[S5]
$y = (y\land \Box x) \lor (y \land \neg \Box x)$

\item[S6]
$\Box (x \lor \Box y ) = \Box x \lor \Box y $

\item[S7]
$\Box (\neg x \lor (y \land x)) \leq \neg \Box x \lor \Box y $

\end{enumerate}
\end{prop}

\begin{proof}
Suppose that $A$ is Boolean saturated. S1), S2) and S3) are
trivial. \hspace{0.2cm} S4) Since $x\land y \leq x$ and $x\land y
\leq y$ then $\Box(x\land y) \leq \Box(x) \land \Box(y)$. For the
converse, $\Box(x) \leq x$ and $\Box(y) \leq y$, thus $\Box(x)
\land \Box(y) \leq \Box(x\land y)$. \hspace{0.2cm} S5) Follows
from Proposition \ref{eqcentro} since $\Box(x) \in Z(A)$.
\hspace{0.2cm} S6) For simplicity let $z = \Box y$. From the
precedent item and taking into account that $z \in Z(L)$ we have
that $\Box(z\lor x) \land \Box(\neg z \lor x) = \Box ((z\lor
x)\land (\neg z \lor x )) = \Box(x)$. Since  $\neg z \leq
\Box(\neg z \lor x)$ then we have that $1 = z \lor \neg z \leq z
\lor \Box(\neg z \lor x)$. Also we have $z\leq \Box(z\lor x)$.
Finally $z\lor \Box(x) = (z\lor \Box(z\lor x)) \land
(z\lor\Box(\neg z\lor x)) = (z\lor \Box(z\lor x)) \land 1 =
\Box(z\lor x)$ i.e. $\Box (x \lor \Box y ) = \Box x \lor \Box y $.
\hspace{0.2cm} S7) Since $\Box(x) \leq x$ then $\neg x \leq  \neg
\Box x$, we have that $\neg x \lor (y \land x) \leq \neg \Box x
\lor y$. Using the precedent item $\Box (\neg x \lor (y \land x))
\leq \Box (\neg \Box x \lor y) = \neg \Box x \lor \Box y$ since
$\neg \Box x \in Z(A)$. \\

For the converse, let $a\in A$ and $\{z \in Z(A) : z\leq a \}$. By
$S1$ and $S5$ it is clear that $\Box a \in \{z \in Z(A) : z\leq a
\}$. We see that $\Box a$ is the upper bound of the set. Let $z\in
Z(A)$ such that $z \leq a$ then $1 = \neg z \lor (a \land z)$.
Using $S2$ and $S7$ we have $1 = \Box 1 = \Box (\neg z \lor
(a\land z)) \leq \neg \Box z \lor a = \neg z \lor a $. Therefore
$z = z \land (\neg z \lor \Box a )$ and since $z$ is central $z =
z \land \Box a$ resulting $z \leq \Box a$. Finally $ \Box a =
Max\{z \in Z(A) : z\leq a \} $.

\end{proof}

\begin{theo}
The class of Boolean saturated orthomodular lattices constitutes a
variety which is axiomatized by

\begin{enumerate}
\item
Axioms of ${\cal OML}$

\item
$S1,...,S7$

\end{enumerate}
\end{theo}

\begin{proof}
Obvious by Proposition \ref{PROST}
\end{proof}

Boolean saturated orthomodular lattices are algebras $ \langle A,
\land, \lor, \neg, \Box, 0, 1  \rangle$ of type $ \langle 2, 2, 1,
1, 0, 0 \rangle$ and the variety they constitute will be noted as
${\cal OML}^\Box$.

On each algebra of ${\cal OML}^\Box$ we can define the possibility
operator as unary operation $\Diamond$ given by

$$\Diamond x = \neg \Box \neg x$$

\begin{prop}\label{POS}
Let $A$ be a Boolean saturated orthomodular lattice and $a, b \in
A$. Then we have

\begin{enumerate}

\item
$a\leq \Diamond a$

\item
$\Diamond a = Min \{z\in Z(A): a\leq z  \}$

\end{enumerate}
\end{prop}

\begin{proof}
We first note that $\Diamond a \in Z(A)$ since  $\Box \neg a \in
Z(A)$. On the other hand $\Box \neg a \leq \neg a$ and then $a =
\neg \neg a \leq \neg \Box \neg a = \Diamond a$. If $z \in Z(A)$
such that $a \leq z$ then $\neg z \leq \neg a$ resulting $\neg z
\leq \Box \neg a$. Thus $\Diamond a = \neg \Box \neg a \leq z$.

\end{proof}

\begin{rema}
{\rm From \ref{POS} it may be seen that Boolean saturated
orthomodular lattices satisfy the four items which motivate our
approach. }
\end{rema}

\begin{theo}\label{COMPZ}
Let $L$ be an orthomodular lattice. Then there exists an
orthomodular monomorphism $f:L \rightarrow L^\Box$ such that
$L^\Box \in {\cal OML}^\Box$.
\end{theo}

\begin{proof}
Let $f: L \rightarrow \prod_{i\in I} L_i$  be a subdirect
embedding of $L$. Since $L_i$ is subdirectly irreducible then
$Z(L_i) = \{0,1\}$ for each $i\in I$ resulting $Z(\prod_{i\in I}
L_i)$ a complete Boolean algebra and so $\prod_{i\in I} L_i$ is
Boolean saturated.
\\
\end{proof}

In this case we may say that each orthomodular lattice can be
represented by a Boolean saturated one. In this sense,  the
embedding of orthomodular systems in modal systems proposed in
\cite{Dish} is maintained. On the other hand, we see that the
defined modal operators are quantifiers in the sense of
\cite{JAN1, JAN2}.

\vskip 0.5truecm

For each orthomodular lattice $L$, if $f:L \rightarrow L^\Box$
such that $L^\Box \in {\cal OML}^\Box$ is an orthomodular
monomorphism, we refer to $L^\Box$ as a {\it modal extension} of
$L$. In this case, we may see the lattice $L$ as a subset of
$L^\Box$.

\section{Modalities: enlargement of the expressivity of the
discourse}

It is clear that the addition of modalities  gives by itself
greater expressive power to the language of propositions about the
system. But what we want to emphasize is that it gives an adequate
framework to represent, for example,  the Born rule for the
probability of actualization of a property, something that  has no
place in the orthomodular lattice alone. In order to develop these
ideas, we need to prove which conditions on elements of a subset
$A$ are necessary to make $\langle A \rangle_L $, the sublattice
generated by $A$, a Boolean sublattice.

\begin{definition}
{\rm Let $L$ be an orthomodular lattice and $a,b \in L$. Then $a$
commutes with $b$ if and only if $a = (a\land b) \lor (a \land
\neg b)$. A non-empty subset $A$ is called a Greechie set iff for
any three different elements of $A$, at least one commutes with
the other two. }
\end{definition}

\begin{prop}\label{GREE}
Let $L$ be an orthomodular lattice. If $A$ is a Greechie set in
$L$ such that for each $a\in A, \neg a \in A$ then, $\langle A
\rangle_L $ is Boolean sublattice.
\end{prop}

\begin{proof}
It is well known from {\rm \cite{GREE}} that $\langle A \rangle_L
$ is a distributive sublattice of $L$. Since distributive
orthomodular lattices are Boolean algebras, we only need to see
that $\langle A \rangle_L $ is closed by $\neg$. To do that we use
induction on the complexity of terms of the subuniverse generated
by  $A$. For $comp(a) = 0$, it follows from the fact that $A$ is
closed by negation. Assume validity for terms of the complexity
less than $n$. Let $\tau$ be a term such that $comp(\tau)= n$. If
$\tau = \neg \tau_1$ then $\neg \tau \in \langle A \rangle_L$
since $\neg \tau = \neg \neg \tau_1 = \tau_1$ and $\tau_1 \in
\langle A \rangle_L$. If $\tau = \tau_1 \land \tau_2$, $\neg \tau
= \neg \tau_1 \lor \neg \tau_2$. Since $comp(\tau_i) < n$, $\neg
\tau_i \in \langle A \rangle_L$ for $i= 1,2$ resulting $\neg \tau
\in \langle A \rangle_L$. We use the same argument in the case
$\tau = \tau_1 \lor \tau_2$. Finally $\langle A \rangle_L$ is a
Boolean sublattice.
\end{proof}

\begin{definition}
{\rm Let $L$ be an orthomodular lattice and $L^\Box \in {\cal
OML}^\Box$ be a modal extension of $L$. We define the {\it
possibility space} of $L$ in  $L^\Box$ as

$$\Diamond L  = \langle \{\Diamond p : p \in L \} \rangle_{L^\Box}
$$

}
\end{definition}

The {\it possibility space} represents the modal content added to
the discourse about properties of the system.

\begin{prop}\label{POSSPACE}
Let $L$ be an orthomodular lattice, $W$  a Boolean sublattice of
$L$ and $L^\Box \in {\cal OML}^\Box$ a modal extension of $L$.
Then $\langle W \cup \Diamond L \rangle_{L^\Box}$ is a Boolean
sublattice of $L^\Box$. In particular $\Diamond L$ is a Boolean
sublattice of $Z(L^\Box)$.

\end{prop}

\begin{proof}
Follows from Proposition \ref{GREE} since $W \cup  \Diamond L$  is
a Greechie set closed by $\neg$.
\\
\end{proof}

We know that, in the orthomodular lattice of the properties of the
system, it is always possible to choose a context in which any
possible property pertaining to this context can be considered as
an actual property. We formalize this fact in the following
definition and then we prove that this is always possible in our
modal structure.

\begin{definition}
{\rm Let $L$ be an orthomodular lattice, $W$ a Boolean sublattice
of $L$, $p\in W$ and $L^\Box$ be a modal extension of $L$. If $f:
\Diamond L \rightarrow {\bf 2}$ is a Boolean homomorphism such
that $f(\Diamond p) = 1$ then an actualization of $p$ compatible
with  $f$ is a Boolean homomorphism $f_p: W \rightarrow {\bf 2}$
such that

\begin{enumerate}

\item
$f_p(p) = 1$

\item
There exists a Boolean homomorphism  $g : \langle W \cup  \Diamond
L \rangle_{L^\Box} \rightarrow {\bf 2}$ such that $g\mid W = f_p$
and $g \mid \Diamond L = f $

\end{enumerate}

}

\end{definition}

\begin{theo} \label{ACT}
Let $L$ be an orthomodular lattice, $W$ a Boolean sublattice of
$L$, $p\in W$ and $L^\Box$ be a modal extension of $L$. If $f:
\Diamond L \rightarrow {\bf 2}$ is a Boolean homomorphism such
that $f(\Diamond p) = 1$ then there exists an actualization of $p$
compatible with $f$.
\end{theo}

\begin{proof}
Let $F$ be the filter associated with the Boolean homomorphism
$f$. We consider the $\langle W \cup  \Diamond L
\rangle_{L^\Box}$-filter $F_p$ generated by $ F \cup \{p \}$. We
want to see that $F_p$ is a proper filter. If $F_p$ is not proper,
then there exists $a\in F$ such that $a \land p \leq 0$. Thus
$p\leq \neg a$ being $\neg a$ a central element. But $\Diamond p$
is the smallest Boolean element greater than $p$, then $\Diamond p
\leq \neg a$ or equivalently $\Diamond p \land a = 0$ And this is
a contradiction since  $\Diamond p, a \in F $ being $F$ a proper
filter. Thus we may extend $F_p$ to be a maximal filter  $F_M$ in
$\langle W \cup  \Diamond L \rangle_{L^\Box}$, resulting the
natural projection $\langle W \cup \Diamond L \rangle_{L^\Box}
\rightarrow \langle W \cup  \Diamond L \rangle_{L^\Box} /F_M
\approx {\bf 2} $ an actualization of $p$ compatible with $f$.

\qed
\\

\end{proof}

The next theorem allows an algebraic representation of the Born
rule which quantifies possibilities from a chosen spectral
algebra.

\begin{theo} \label{BORN}
Let $L$ be an orthomodular lattice, $W$ a Boolean sublattice of
$L$ and $f: W \rightarrow {\bf 2}$  a Boolean homomorphism. If we
consider a modal extension $L^\Box$ of $L$ then there exists a
Boolean homomorphism $f^*: \langle W \cup  \Diamond L
\rangle_{L^\Box} \rightarrow {\bf 2} $ such that $f^* \mid W = f$.

\end{theo}

\begin{proof}
Let $i: W \rightarrow \langle W \cup  \Diamond L \rangle_{L^\Box}$
be the Boolean canonical embedding. If we consider the following
diagram:

\begin{center}
\unitlength=1mm
\begin{picture}(60,20)(0,0)
\put(8,16){\vector(3,0){5}} \put(2,10){\vector(0,-2){5}}

\put(2,16){\makebox(0,0){$W$}} \put(20,16){\makebox(0,0){$\bf 2$}}
\put(2,0){\makebox(0,0){$ \langle W \cup  \Diamond L
\rangle_{L^\Box} $}}

\put(2,20){\makebox(17,0){$f$}} \put(2,8){\makebox(-5,0){$i$}}
\end{picture}
\end{center}

\noindent we see that there exists a Boolean homomorphism $f^*:
\langle W \cup \Diamond L \rangle_{L^\Box} \rightarrow {\bf 2}$
such that $f^* \mid W_A = f$ because ${\bf 2}$ is injective in the
variety of Boolean algebras \cite{SIK}.
\\
\end{proof}

\noindent We note that this reading of the Born rule is a  kind of
converse of the possibility of actualizing properties given by
 Theorem \ref{ACT}.

\section{Kochen-Specker theorem: a limit also for modalities}

The addition of modalities to the discourse about the properties
of a quantum system enlarges its expressive power. At first sight
it may be thought that this could help to circumvent
contextuality, allowing to refer to physical properties belonging
to the system in an objective way that resembles the classical
picture. But this is not the case as we have announced in
\cite{DFR}. To prove it here, we introduce an algebraic
representation of the notion of  global actualization:

\begin{definition}

{\rm Let $L$ be an orthomodular lattice, $(W_i)_{i \in I}$ the
family of Boolean sublattices of $L$ and $L^\Box$ a modal
extension of $L$. If $f: \Diamond L \rightarrow {\bf 2}$ is a
Boolean homomorphism, an {\it actualization compatible with } $f$
is a global valuation $(v_i: W_i \rightarrow {\bf 2})_{i\in I}$
such that $v_i\mid W_i \cap \Diamond L = f\mid W_i \cap \Diamond L
$ for each $i\in I$.}
\end{definition}
Compatible actualizations represent the passage from possibility
to actuality.

\begin{theo}
Let $L$ be an orthomodular lattice. Then $L$ admits a global
valuation iff for each possibility space there exists a Boolean
homomorphism  $f: \Diamond L \rightarrow {\bf 2}$ that admits  a
compatible actualization.
\end{theo}

\begin{proof}
Suppose that $L$ admits a global valuation $(v_i: W_i \rightarrow
{\bf 2})_{i\in I}$. Let ${L^\Box}$ be a modal extension of $L$ and
consider $A_i = W_i \cap \Diamond L$. Let $f_0 = \bigcup_i A_i
\rightarrow {\bf 2}$ such that $f_0(x) = v_i(x)$ if $x\in W_i$.
$f_0$ is well defined since $(v_i)_i$ is a global valuation. If we
consider $\langle \bigcup_i A_i \rangle_{L^\Box}$, the subalgebra
of $L^{\Box}$ generated by the join of the family $(A_{i})$, it
may be proved that it is a Boolean subalgebra of the possibility
space $\Diamond L$. We can extended $f_0$ to a Boolean
homomorphism $f_0^*:\langle \bigcup_i A_i \rangle_{L^\Box}
\rightarrow {\bf 2}$. Since ${\bf 2}$ is injective in the variety
of Boolean algebras \cite{SIK}, there exits a Boolean homomorphism
$f: \Diamond L \rightarrow {\bf 2}$ such that the following
diagram is commutative

\begin{center}
\unitlength=1mm
\begin{picture}(20,20)(0,0)
\put(12,16){\vector(3,0){8}} \put(2,10){\vector(0,-2){5}}
\put(10,4){\vector(1,1){7}}

\put(2,10){\makebox(13,0){$\equiv$}}

\put(2,16){\makebox(0,0){$\langle \bigcup_i A_i \rangle_{L^\Box}
$}} \put(23,16){\makebox(0,0){${\bf 2}$}}
\put(2,0){\makebox(0,0){$\Diamond L$}}
\put(7,20){\makebox(17,0){$f_0^*$}} \put(2,8){\makebox(-6,0){$i$}}
\put(18,2){\makebox(-4,3){$f$}}
\end{picture}
\end{center}

Thus $f: \Diamond L \rightarrow {\bf 2}$ admits  a compatible
actualization. The converse is immediate.
\\
\end{proof}

Since the possibility space is a Boolean algebra, there exists a
Boolean valuation of the possible properties. But in view of the
last theorem, an actualization that would correspond to a family
of compatible global valuations is prohibited. Thus the theorem
states that the contextual character of quantum mechanics is
maintained even when the discourse is enriched with modalities.

\section{Conclusions}

From the algebraic characterization of contextuality given by the
non existence of compatible global valuations over the
orthomodular structure, we show that, if this structure is
enriched with modal operators, the discourse about properties is
genuinely enlarged. However, the  contextual character of the
complete language is maintained. Thus contextuality remains a main
feature of quantum systems even when modalities are taken into
account.

\section*{Acknowledgements}
 This work was partially supported by the following grants: PICT
04-17687 (ANPCyT), PIP N$^o$ 1478/01 (CONICET), UBACyT N$^o$ X081
and X204. G. D. is fellow of the Consejo Nacional de
Investigaciones Cient\'{\i}ficas y T\'ecnicas (CONICET).

\end{document}